\documentclass[%
 reprint,
superscriptaddress,
 amsmath,amssymb,
]{revtex4-2}

\usepackage{comment}
\usepackage{multirow}

\usepackage{amsthm}  

\theoremstyle{plain}
\newtheorem{theorem}{Theorem}

\usepackage{graphicx}
\usepackage{dcolumn}
\usepackage{verbatim} 
\usepackage{bm}
\usepackage[hidelinks]{hyperref}
\usepackage{pdfpages}

\usepackage{pgffor} 

\makeatletter
\AtBeginDocument{\let\LS@rot\@undefined}
\makeatother

\def\supplementfilename{supp-optical_antimatter_final.pdf}

\pdfximage{\supplementfilename}
\def\numbersupplementpages{\the\pdflastximagepages}

\newif\ifarXiv
\arXivtrue

\usepackage[table]{xcolor}
\definecolor{red}{rgb}{1,0,0}
\def\OL#1{{\color{red} [#1]}} 
\usepackage{comment}

\usepackage{textcomp}

\begin{document}

\title{Ideal Optical Antimatter using Passive Lossy Materials \\ 
under Complex Frequency Excitation}%

\author{Olivia Y. Long}
\email{olong@stanford.edu}
 \affiliation{Department of Applied Physics, Stanford University, Stanford, California 94305, USA}
 \affiliation{Edward L. Ginzton Laboratory, Stanford University, Stanford, California 94305, USA}

 \author{Peter B. Catrysse}
 \affiliation{Edward L. Ginzton Laboratory, Stanford University, Stanford, California 94305, USA}

\author{Seunghoon Han}
\affiliation{Samsung Advanced Institute of Technology, Samsung Electronics, Suwon-si 16678, South Korea}
\affiliation{Semiconductor R\&D Center, Samsung Electronics, Hwaseong-si 18448, South Korea}

\author{Shanhui Fan}

\email{shanhui@stanford.edu}
\affiliation{Department of Applied Physics, Stanford University, Stanford, California 94305, USA}
\affiliation{Edward L. Ginzton Laboratory, Stanford University, Stanford, California 94305, USA}
\affiliation{Department of Electrical Engineering,
Stanford University, Stanford, California 94305, USA}

\date{\today}

\begin{abstract}

The original concept of left-handed material has inspired the possibility of optical antimatter, where the effect of light propagation through a medium can be completely cancelled by its complementary medium. Despite recent progress in the development of negative-index metamaterials, losses continue to be a significant barrier to realizing optical antimatter. 
In this work, we show that passive, lossy materials can be used to realize optical antimatter when illuminated by light at a complex frequency. We further establish that one can engineer arbitrary complex-valued permittivity and permeability in such materials. Strikingly, we show that materials with a positive index at real frequencies can act as negative-index materials under complex frequency excitation. Using our approach, we numerically demonstrate the optical antimatter functionality, as well as double focusing by an ideal perfect lens and superscattering.
 Our work demonstrates the power of temporally structured light in unlocking the promising opportunities of complementary media, which have until now been inhibited by material loss.


\end{abstract}

\maketitle

\section{Introduction}

An intriguing effect enabled by the developments of metamaterials is the possibility to create optical antimatter or complementary media. The electromagnetic scattering of a medium with an inhomogeneous distribution of permittivity $\epsilon$ and permeability $\mu$ can be completely eliminated, if one places a medium with complementary permittivity and permeability distribution $-\epsilon, -\mu$ next to it \cite{focusing_light_using_neg_refrac_2003}. The notion of optical antimatter is a generalization of the effect of perfect lensing with negative refraction \cite{pendry_prl_2000, veselago_1968}, and has further implications for transformation optics, ranging from superscattering \cite{superscatterer_pendry_2009, superscat_cloak_nanowires_2013, zhichao_superscattering_PRL_2010, fan_superscat_CMT_2012} to perfect invisibility devices \cite{Leonhardt_2006_gen_relativity_EE, pendry_transformation_optics_2006, schurig_microwave_cloak_2006, optical_cloaking_2007, dielectric_optical_cloak_2009}, which generally require spatially inhomogeneous materials. 

For a regular material with a positive permittivity and permeability, its complementary medium is a negative index medium with negative permittivity and permeability. However, due to the constraint of a positive energy density, negative permittivity or permeability necessarily implies that the material is dispersive \cite{smith_kroll_NIM_need_dispersion_2000}. By the Kramers-Kronig relations then, such a material must also be lossy. In experiments, the presence of loss has severely limited the demonstration of such complementary media \cite{near-sighted_superlens_2005, shelby_exp_neg_index_2001,nanofab_neg_perm_visible_freq_2005, Shalaev_2005_high_NIM_loss_visible, three_dim_photonic_metamat_optic_freq_2008, soukoulis_neg_refrac_optical_review_2007, zheludev_metamaterial_review_2010, smith_limitations_subdiffrac_imaging_2003}. 

In recent years, there have been emerging interests in considering the electromagnetic properties of materials at complex frequencies \cite{complex_freq_photonics_alu_science_2025, lalanne_quasinormal_modes_2018}. In particular, it has been theoretically proposed and experimentally demonstrated that the use of complex frequencies can significantly improve the resolution of negative refraction lens where the material has intrinsic losses \cite{overcoming_losses_superlenses_science_2023, loss_compensation_alu_PRX_2023, superlens_time_domain_greffet_2012, time-driven_superoscillation_neg_refrac}. Many other effects associated with the use of complex frequency excitations, such as coherent virtual absorption \cite{coherent_virtual_absorption_2017},
virtual parity-time symmetry \cite{virtual_PT_symm_2020},
 polariton propagation \cite{compensating_losses_polariton_2024}, light stopping \cite{dispersionlesss_light_waveguides_PRL_2014}, virtual critical coupling \cite{virtual_crit_coupling_2024, virtual_crit_coupling_theory_2020}, and transient non-Hermitian skin effect \cite{transient_nhse_2022} have also been considered. Building upon these works, here we show that a pair of complementary media can be constructed from only \textit{lossy} materials with the use of complex frequency excitations. 
 Remarkably, we show that such complementary media can be achieved using only \textit{positive-index} dispersive, passive materials. In doing so, we also establish the possibility of engineering arbitrary complex-valued permittivity and permeability in such materials. We further demonstrate that our approach can be applied to realize optical antimatter, double refocusing, and superscattering.
 Our work opens an avenue to the experimental exploration of optical antimatter by overcoming the effects of loss that is intrinsic to these systems.

\section{Results}
\subsection{Theory}

We start by briefly reviewing the concept of complementary media, as originally proposed in Ref. \cite{focusing_light_using_neg_refrac_2003}. For an isotropic medium with an electric permittivity $\epsilon$ and a magnetic permeability $\mu$, its complementary medium has an electric permittivity $\tilde \epsilon$ and a magnetic permeability $\tilde \mu$ that satisfies:
\begin{align}
\tilde \epsilon = -\epsilon, \tilde \mu = - \mu \label{complement_defn}
\end{align}
Moreover, Ref. \cite{focusing_light_using_neg_refrac_2003} envisioned an ideal scenario where the two complementary media are both lossless. Here in this section, we show that such an ideal complementary pair can be realized at a single complex frequency using lossy, passive materials. 

\subsubsection{Lossless Propagation in Complementary Media using Lossy, Passive Materials at Complex Frequency}
We first discuss the condition for lossless propagation of light at a complex frequency. 
This requires the wavevector $k$ to be purely real. 
Using the $e^{-i \omega t}$ time convention, Maxwell's equations state:
\begin{align}
    k \times E &= \omega \mu(\omega) H \\
    k \times H &= -\omega \epsilon(\omega) E
\end{align}
where we have explicitly included the frequency dependence of $\epsilon$ and $\mu$. 
When $\omega \in \mathbb{C}$, the following conditions
\begin{align}
    \omega \epsilon(\omega) &\in \mathbb{R} \label{omega_eps_real_cond}\\
    \omega \mu(\omega) &\in \mathbb{R} \label{omega_mu_real_cond}
\end{align}
are sufficient in order to have $k \in \mathbb{R}$. 
%
Importantly, at a complex frequency, to achieve lossless propagation of light, the permittivity and permeability must also be complex.

We now proceed to show that a 
complementary media pair with lossless propagation in both media can be achieved using lossy, passive materials excited at a complex frequency. For this purpose, we consider the Lorentz-Drude model, which is both a fundamental model for a dispersive material and a common model used in the design of metamaterials \cite{effec_permittiv_permeab_metamaterial_PRB_2002}:
\begin{align}
    \epsilon( \omega) = 1 + \frac{\omega_p^2}{\omega_0^2 - \omega^2 - i\omega \gamma} \label{LD_model}
\end{align}
In this model, $\omega_0$ is the frequency of the material resonance, and hence:
\begin{align}
    \omega_0^2 > 0 \label{omega0_cond}.
\end{align}
The plasma frequency is $\omega_p$ and passivity requires that \cite{siegman1986lasers}
\begin{align}
\omega_p^2 > 0 \label{omegap_cond}
\end{align}
while causality requires the damping rate $\gamma$ to satisfy \cite{Landau_electrodyn_continuous_media_1984}
\begin{align}
    \gamma > 0 .\label{gamma_cond}
\end{align}

Our argument is based on the following theorem.
\begin{theorem}
 Given a complex frequency $\omega = \omega^\prime + i\omega^{\prime \prime}$ with $\omega^{\prime \prime} < 0$, and any complex value $C + iD$ where $C,D\in \mathbb{R}$, there exist parameters $\omega_0^2, \omega_p^2, \gamma$ satisfying Eqs. \eqref{omega0_cond}--\eqref{gamma_cond} such that $\epsilon(\omega) = C+ iD$. 
\end{theorem}

We now prove this theorem.  
Equating the Lorentz-Drude expression in Eq. \eqref{LD_model} to 
$C+iD$ and matching the real and imaginary parts yields a system of equations described by 
\begin{align}
    M\mathbf{x} = \mathbf{b} \label{matrix_eqn}
\end{align} where \cite{SM}:
\begin{align}
    M &= { \setlength{\arraycolsep}{9pt} 
 \begin{bmatrix}
        C-1 & 1 & \omega^{\prime \prime}(C-1) - \omega^\prime D \\
        D & 0 & \omega^\prime (C-1) + \omega^{\prime \prime} D
    \end{bmatrix}
    } \\
    \mathbf{x} &=
    \begin{bmatrix}
        \omega_0^2 \\ \omega_p^2 \\ \gamma
    \end{bmatrix} \\
    \mathbf{b} &= \begin{bmatrix}
        (C-1)(\omega^{\prime 2} - \omega^{\prime \prime ^2}) + 2\omega^\prime \omega^{\prime \prime}D \\
        D(\omega^{\prime 2}-\omega^{\prime \prime 2}) - 2\omega^\prime \omega^{\prime \prime}(C-1)
    \end{bmatrix} \label{eps_matrix_eqn}
\end{align}
We thus wish to show that there always exists a solution $\mathbf{x}$ with all components of $\mathbf{x}$ being positive for arbitrary $C, D \in \mathbb{R}$. 

We first identify a particular solution: 
\begin{align}
    \mathbf{x}_p = \begin{bmatrix}
    \omega^{\prime 2} + \omega^{\prime \prime 2}
    \\
    0 \\
    - 2\omega^{\prime \prime}
    \end{bmatrix}
\end{align}
Since we have set $\omega^{\prime \prime} < 0$, the nonzero components of $\mathbf{x}_p$, i.e. $\omega^{\prime 2} + \omega^{\prime \prime 2},- 2\omega^{\prime \prime} > 0$. 

%
Since the system of equations in Eq. \eqref{matrix_eqn} is underdetermined, the complete solution $\mathbf{x}$ is:
\begin{align}
    \mathbf{x} = \mathbf{x}_p + \alpha \mathbf{x}_n \label{complete_x_soln}
\end{align}
where 
$\alpha$ is a free variable and
\begin{align}
    \mathbf{x}_n &= \mathbf{r}_2 \times \mathbf{r}_1 \nonumber \\
    &= \begin{bmatrix}
        \omega^\prime (1-C) - \omega^{\prime \prime } D \\
        \omega^\prime\Big( D^2 + (C-1)^2 \Big)
        \\
        D
    \end{bmatrix}
\end{align} 
where $\mathbf{r}_1$ and $\mathbf{r}_2$ are the first and the second rows of $M$, respectively. 
Eq. \eqref{complete_x_soln} can be derived noting that $M \mathbf{x}_n = \mathbf{0}$.
We now perform a component-wise analysis. Let $x_k$ denote the $k$th component of $\mathbf{x}$. From Eq. \eqref{complete_x_soln}, each component can thus be written as:
\begin{align}
x_k(\alpha) = (x_p)_k + \alpha\,(x_n)_k \label{component_eqn}
\end{align} 
Without loss of generality, we can assume $\omega^\prime > 0$. In this case, $x_2(\alpha) > 0$ for any choice of $\alpha > 0$. On the other hand, since $(x_p)_1$ and $(x_p)_3$ are both positive, with a sufficiently small $\alpha > 0$, by continuity, both $x_1(\alpha)$ and $x_3(\alpha)$ remain positive. Therefore, there exists a solution $\mathbf{x}$ with all its components being positive. This concludes our proof of the theorem. With a bit of further algebra, the procedure above in fact can be used to determine the range of $\alpha$ in which the components of $\mathbf{x}$ are all positive \cite{SM}. 
The theorem above can likewise be applied to $\mu(\omega)$, which can also be described by a similar Lorentz-Drude model \cite{effec_permittiv_permeab_metamaterial_PRB_2002}. 

In metamaterials, the parameters $\omega_p^2$, $\omega_0^2$ and $\gamma$ can all be controlled: $\omega_p^2$ is related to the density of meta-atoms, $\omega_0^2$ is controlled by the geometry of an individual meta-atom, and $\gamma$ is related to the choice of constituent materials. Our theorem thus indicates substantial new flexibility in achieving arbitrary electromagnetic response, using lossy passive metamaterials, by operating at complex frequencies. 

We apply the theorem above to the design of optical antimatter. 
At a given complex frequency $\omega$ with $\omega^{\prime \prime}<0$,  we can choose two arbitrary real positive numbers $q_\epsilon$ and $q_\mu$. From our theorem, there exists a pair of lossy, passive Lorentz-Drude materials as described by $\epsilon(\omega)$, $\mu(\omega)$ and $\tilde \epsilon(\omega)$, $\tilde \mu (\omega)$, respectively, such that
\begin{align}
\tilde \epsilon(\omega) &= - \epsilon(\omega) = q_\epsilon/\omega \\ \tilde \mu(\omega) &= -\mu(\omega) = q_\mu/\omega
\end{align}
These materials form a complementary pair with both supporting lossless propagation. 

\begin{figure*}[!]
\centering
{
\includegraphics[width=0.95\textwidth]{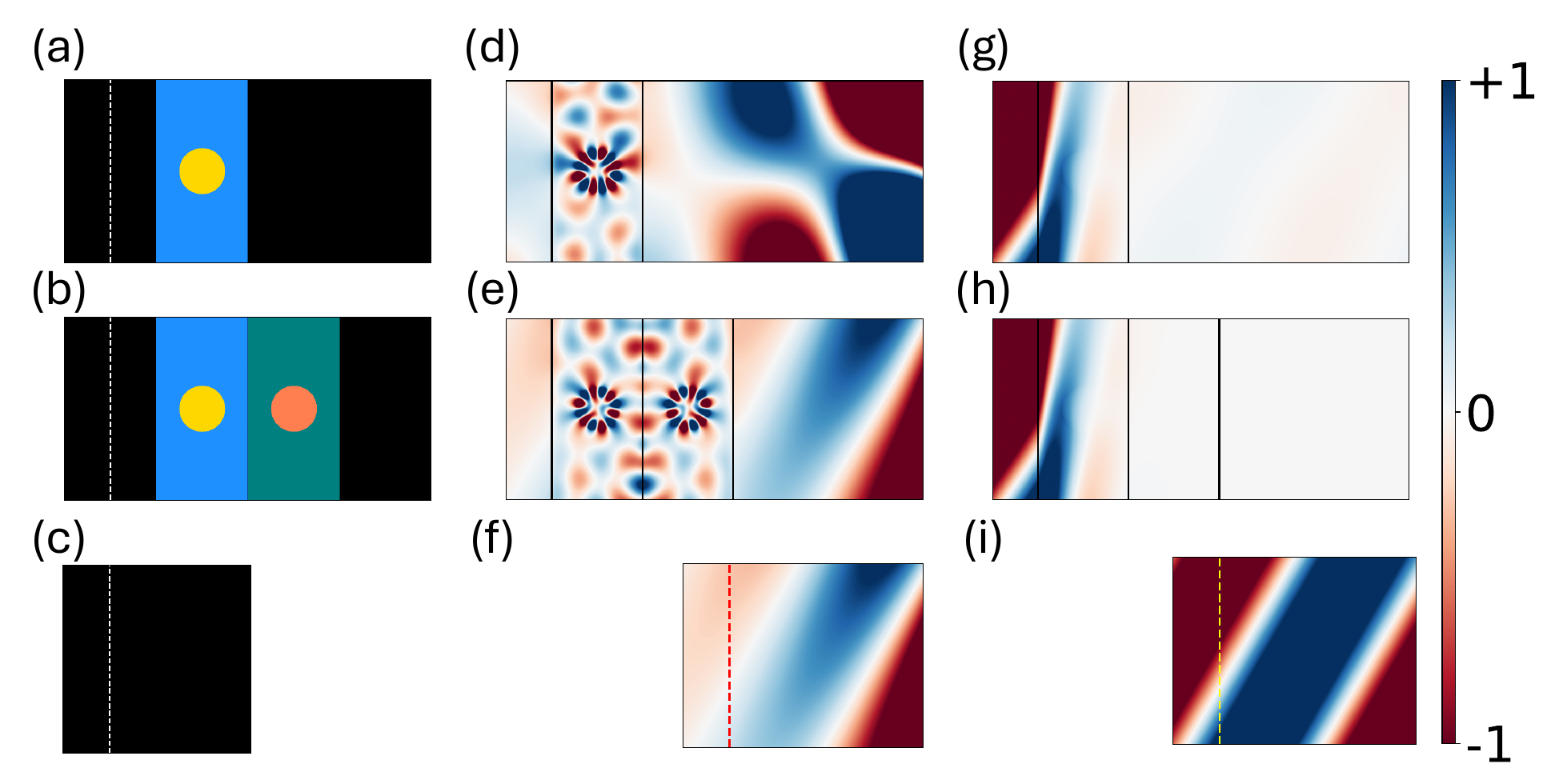}
}
\caption{Optical antimatter at complex frequency. 
In panels (a)--(c), the colors blue, teal, yellow and orange correspond to Media 1, 2, 3, 4, as described in Table \ref{lorentz_param_table}, respectively. Black denotes vacuum.
(a) Schematic of initial structure with circular scatterer of radius 150 nm embedded in slab of thickness 0.6 $\mu$m. Dashed white line indicates location of TM-polarized plane wave source. (b) Schematic of structure placed adjacent to its complementary counterpart.
(c) Free space configuration with plane wave source. 
(d)--(f) TM-polarized Re[$H_z$] field patterns at the complex frequency $\omega = \omega_c$ for the configurations in (a), (b), and (c), respectively. (g)--(i) TM-polarized Re[$H_z$] field patterns at the real frequency $\omega = \omega_r$ for the configurations in (a), (b), and (c), respectively. In (f) and (i), the red and yellow dashed lines mark the effective location of the optically cancelled complementary media pair, respectively. For each $\omega$, field plots are normalized to the same maximum and minimum values. 
 All field plots were generated using FDFD \cite{ceviche_ref_2019}, with Bloch boundary conditions in the transverse ($y$) direction and PML boundary conditions in the propagation ($x$) direction. Plots show 0th diffraction order in free space. 
}
\label{optical_antimatter}
\end{figure*}

\subsubsection{Achieving Negative Index Behavior in Positive Index Media at Complex Frequency}
As can be seen in Eq. \eqref{complement_defn}, complementary media are closely related to negative index media. As a side note, we observe that a medium that has a positive index at a real frequency may possess a negative index 
at a complex frequency with the same real component. To illustrate this point, we consider a medium with a permittivity $\epsilon(\omega)$ and a permeability $\mu_0$. 
%
%
%
%
Without loss of generality, we consider TM-polarized waves propagating in the $z$-direction:
\begin{align}
    \mathbf{E}(z) &= e^{i k z}  (E_x, 0, 0) \\
    \mathbf{H}(z) &= e^{i k z}  (0, H_y, 0) 
\end{align}
Expressing the refractive index $n$ in terms of its real and imaginary components: $n=n^\prime + i n^{\prime \prime}$. The wavevector $k$ can then be written as:
\begin{align}
   k = \frac{\omega n}{c} &= \frac{(\omega^\prime + i\omega^{\prime \prime}) (n^\prime + i n^{\prime \prime}) }{c} \nonumber \\
   &= \frac{(\omega^\prime n^\prime - \omega^{\prime \prime}n^{\prime \prime}) + i(\omega^\prime n^{\prime \prime} + \omega^{\prime \prime} n^\prime)}{c}
\end{align}
Thus, the direction of the phase velocity is controlled by the sign of $\text{Re}[k] = \omega^\prime n^\prime - \omega^{\prime \prime}n^{\prime \prime}$.
On the other hand, the direction of the power flow is determined by the sign of the $z$-component of the time-averaged Poynting vector $\langle \mathbf{S} \rangle = \frac{1}{2}\text{Re}[\mathbf{E} \times \mathbf{H}^*]$: 
\begin{align}
    \langle S_z \rangle = \frac{|E_x|^2}{2} \sqrt{\frac{\epsilon_0}{\mu_0}}n^\prime e^{-2(\omega^\prime n^{\prime \prime} + \omega^{\prime \prime} n^\prime)/c}
\end{align}
Here, the sign of $n^\prime$ 
determines the direction of $\langle S_z \rangle$. 
When $\omega$ is real, i.e. $\omega^{\prime \prime}=0$, the phase and the group velocity have the same sign, as expected since this material has a magnetic permeability of $\mu_0$
\cite{McCall_neg_index_demystified_2002}. However, for a complex $\omega$, it becomes possible to have $\omega^{\prime \prime}n^{\prime \prime} > \omega^\prime n^\prime $, while keeping the sign of $n^\prime$ to be positive. 
Thus, negative index behavior can be achieved at 
the complex $\omega$. 

Having established the theoretical basis for complementary media with lossless propagation using only passive, lossy materials at a complex frequency, we next illustrate the optical antimatter functionality with numerical simulations.
Although there has been an experimental work on optical antimatter for air, the impedance mismatch between air and the structure prevents the full recovery of all transverse wave components \cite{exp_demo_optical_antimatter_2009}. Moreover, there have not been any works to date on optical antimatter for materials that are inhomogeneous in the transverse direction. Such inhomogeneity would pave the way to a broader class of applications beyond lensing, including superscattering and cloaking.

\begin{figure*}[t]
\centering
{
\includegraphics[width=0.8\textwidth]{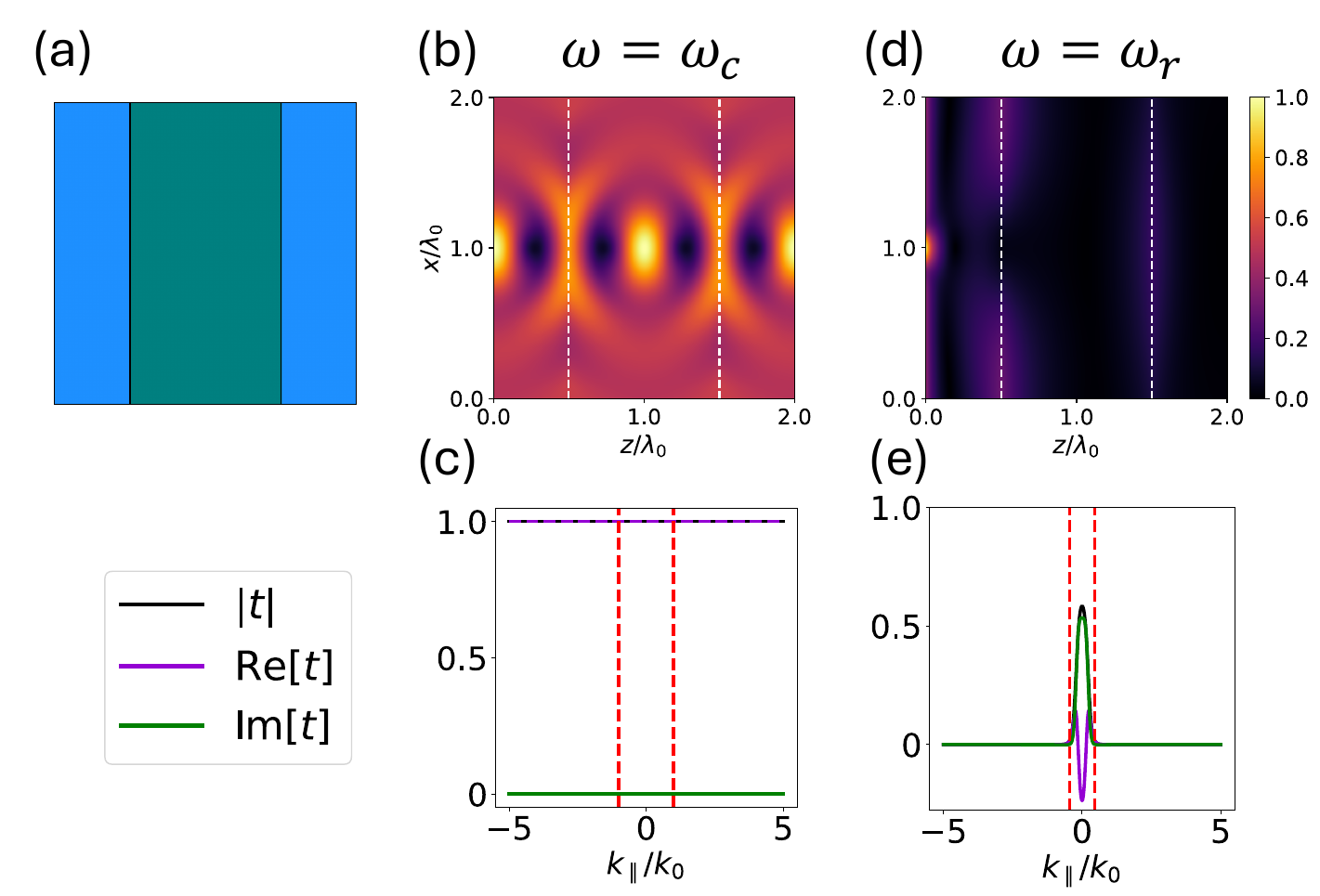}
}

\caption{ Double focusing effect of an ideal perfect lens enabled by optical antimatter under complex frequency excitation. (a) Schematic of three layers comprised of Medium 1, Medium 2, Medium 1 depicted using color scheme of Fig. \ref{optical_antimatter}. Layer thicknesses from left to right: $d_0 = 0.5\lambda_0$, $d_1 = \lambda_0$, $d_2 = 0.5\lambda_0$ where $\lambda_0$ is the wavelength in Medium 1 at $\omega=\omega_c$.  
(b) Magnitude of Poynting vector for a Gaussian point source originating in leftmost layer at complex frequency $\omega = \omega_c$ and at (d) real frequency $\omega = \omega_r$ normalized to their respective maximum values. Dashed white lines indicate interfaces between Medium 1 and Medium 2. For clarity, only propagating waves are plotted in the complex frequency case.
(c) Transmission coefficient through setup shown in panel (a) at $\omega=\omega_c$ and at (e) $\omega=\omega_r$. In both plots, $k_0$ refers to the wavevector defined in Medium 1 at $\omega=\omega_c$. Red dashed lines mark the boundary between evanescent and propagating waves.
All plots shown are for TM polarization.
}
\label{fig_refocus}
\end{figure*}

\begin{figure*}[t]
\centering
{
\includegraphics[width=0.8\textwidth]{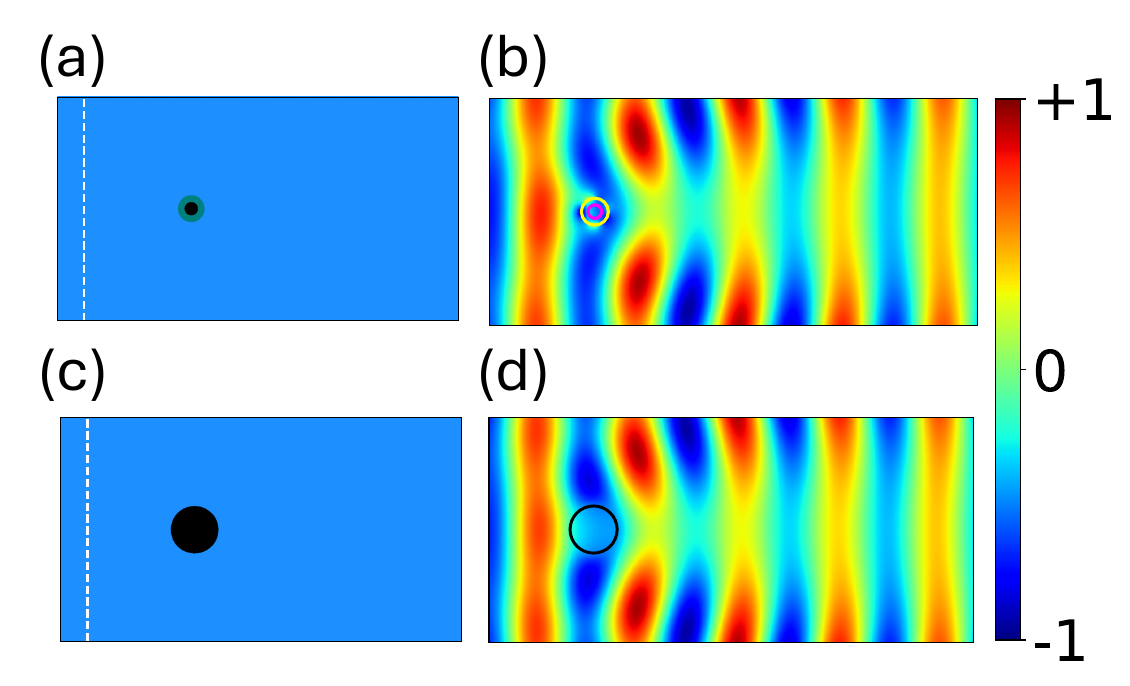}
}
\caption{Superscattering using optical antimatter at complex frequency. (a) Schematic of air scatterer with radius $r=18.75$ nm embedded in Medium 1 and surrounded by an annulus of complementary Medium 2 with thickness $18.75$ nm. (b) Field distribution of Re[$H_z$] in panel (a) under plane wave excitation at complex frequency from left. The yellow circle indicates the boundary between the annulus and the surrounding medium. The magenta circle indicates the boundary of the inner air scatterer. (c) Schematic of air scatterer with radius $R=66.6$ nm embedded in Medium 1. (d) Field distribution of Re[$H_z$] in panel (c) under plane wave excitation at complex frequency from left. The black circle indicates the boundary of the air scatterer.
}
\label{fig_superscat}
\end{figure*}

\subsection{Numerical demonstrations}
\subsubsection{Optical Antimatter}

In our numerical demonstrations, we employ a temporally decaying wave with complex frequency $\omega = \omega_c = (1.256 \times 10^{15} - 4 \times 10^{14}i)$ rad/s. Note that $\text{Im}[\omega_c] < 0$,  consistent with our theory. The carrier frequency $\omega_r = \text{Re}[\omega_c]$ corresponds to light with a free-space wavelength of 1.5 $\mu$m. 
We use two pairs of complementary media described by passive, lossy Lorentz-Drude models. The parameters $\omega_p^2, \omega_0^2, \gamma$, as well as the values of $\epsilon$ and  $\mu$ for these media at frequencies $\omega = \omega_c$ and $\omega = \omega_r$, are provided in Table \ref{lorentz_param_table}. 
\onecolumngrid
\vspace{0.4cm}

\begin{table}[h!]
    \centering
    \begin{tabular}{|c|c|c|c|c|c|c|}
        \hline
        Medium & Parameter & $\omega = \omega_c$ & $\omega = \omega_r$ & $\omega_p$ ($10^{15}$ rad/s) & $\omega_0$ ($10^{15}$ rad/s) & $\gamma$ ($10^{15}$ rad/s) \\ \hline
        \multirow{2}{*}{1} & $\epsilon_1$ & $4+1.274i$ & $3.289+1.292i$ & $3.873$ & $2.559$ & $2.233$ \\ \cline{2-7}
         & $\mu_1$ & $6+1.911i$  & $1.776+1.513i$ & $1.483$ & $1.472$ & $0.917$ \\ \hline
        \multirow{2}{*}{2} & $\epsilon_2$ & $-4-1.274i$ & $0.756 + 1.570i$ & $1.225$ & $1.197$ & $0.743$ \\ \cline{2-7}
        & $\mu_2$ & $-6-1.911i$ & $0.890 + 1.627i$ & $1.245$ & $1.230$ & $0.755$ \\ \hline
        \multirow{2}{*}{3} & $\epsilon_3$ & $12 + 3.822i$ & $2.812+3.399i$ & 2.236 & 1.479 & 0.912 \\ \cline{2-7}
        & $\mu_3$ & $15+4.778i$ & $1.474+1.682i$ & $1.378$ & $1.368$ & $0.833$ \\ \hline
        \multirow{2}{*}{4} & $\epsilon_4$ & $-12 - 3.822i$ & $1.051+1.752i$ & 1.304 & 1.267 & 0.772 \\ \cline{2-7}
        & $\mu_4$ & $-15-4.778i$ & $1.097+1.684i$ & $1.284$ & $1.278$ & $0.777$ \\ \hline
    \end{tabular}
    \caption{Material parameters of Media $1$–$4$ used in numerical demonstrations. $\omega_c = (1.256 \times 10^{15} - 4 \times 10^{14}i)$ rad/s and $\omega_r = \text{Re}[\omega_c] = 1.256 \times 10^{15}$ rad/s.}
    \label{lorentz_param_table}
\end{table}
%
\twocolumngrid
We note that the resonant response in the magnetic permeability, as described by the Lorentz-Drude model, has been observed in optical frequencies using meta-atoms that support magnetic dipole resonances \cite{mag_response_meta_100_Thz_2004, saturation_mag_response_optic_freq_2005, kuznetsovMagneticLight2012}. Similar resonant response in the electric permittivity can be achieved with the use of systems exhibiting resonant electric dipole response such as dielectric meta-atoms \cite{optic_response_si-nano_arrays_prb_2010, staudeTailoringDirectionalScattering2013}. These systems have not been previously used to demonstrate optical antimatter due to the substantial loss. The use of complex frequencies as we propose here overcomes the challenges associated with the loss.

The first complementary pair is comprised of Medium 1 and 2. At $\omega = \omega_c$, $\epsilon_2 = -\epsilon_1$ and $\mu_2 = -\mu_1$.
The second pair is comprised of Medium 3 and 4, where similarly at $\omega = \omega _c$, $\epsilon_4 = -\epsilon_3$ and $\mu_4 = -\mu_3$. At $\omega = \omega_c$, these materials satisfy Eqns. \eqref{omega_eps_real_cond}, \eqref{omega_mu_real_cond} such that the wavevector is purely real and there is lossless propagation of light in each media. 

In Fig. \ref{optical_antimatter}, we showcase the functionality of optical antimatter. Fig. \ref{optical_antimatter}(a) depicts the first structure, where a circular scatterer comprised of Medium 3 is embedded in a slab of Medium 1 that is surrounded by free space. The slab has a thickness of $0.6$ $\mu$m and the scatterer has a radius of $150$ nm. 
The structure is excited by a TM-polarized plane wave at $\omega = \omega_c$, launched from a line source in free space [dashed white line in Fig. \ref{optical_antimatter}(a)] and incident at an angle of 30$^\circ$ with respect to the horizontal direction.
For reference, the same source and the corresponding field pattern in free space are shown in Figs. \ref{optical_antimatter}(c) and (f), respectively. Note that the fields in free space are spatially growing along the propagation direction due to the complex frequency excitation.

%
%
%
When the incident plane wave is transmitted through the structure in Fig. \ref{optical_antimatter}(a), the field pattern that emerges on both sides of the slab, as shown in Fig. \ref{optical_antimatter}(d), is markedly different from that of the plane wave excitation indicating the presence of strong reflection and scattering.  

To demonstrate optical antimatter, we now place the complementary structure adjacent to the first structure, as depicted schematically in Fig. \ref{optical_antimatter}(b). 
Remarkably, in Fig. \ref{optical_antimatter}(e), we see that 
the field patterns outside the combined structure on the left and right sides match precisely to those in free space on the left and right sides of the red line in Fig. \ref{optical_antimatter}(f), respectively. The presence of the complementary structure thus cancels all optical effects of the first structure, rendering the combined structure invisible to the incident light. We also note that the field pattern inside the combined structure has a mirror symmetry with respect to the interface between the first structure and its complementary, even though the field is incident from the left. Such a mirror symmetry is expected for such complementary media, as proven in Ref.  \cite{focusing_light_using_neg_refrac_2003}.

%



We next excite the structures with a TM-polarized plane wave at the corresponding real frequency $\omega=\omega_r$, with the field pattern in free space shown in Fig. \ref{optical_antimatter}(i). As expected, the excitation field propagates in free space without growth or decay along the propagation direction. At $\omega=\omega_r$, the relative permittivity and permeability values for all four media are shown in Table \ref{lorentz_param_table}.  
Notably, all refractive indices are positive. Moreover, all media are lossy, as expected for such passive, lossy Lorentz-Drude models  \cite{Landau_electrodyn_continuous_media_1984}. 
The resulting field patterns for the first structure and for the combined structure are shown in Figs. \ref{optical_antimatter}(g) and (h), respectively. We see that in both cases, the field patterns outside the structure deviate significantly from those of free space. Thus, at this real frequency, the combined structure no longer exhibit the effects of optical antimatter. For additional optical antimatter demonstrations using different geometries, see Supplementary Materials \cite{SM}.

\subsubsection{Ideal Perfect Lens}
One application of complementary media is for the construction of a perfect lens \cite{pendry_prl_2000}. As a second illustration, we show that an ideal perfect lens can be achieved using passive, lossy Lorentz-Drude materials excited at complex frequency.  
%
%
%
We consider three planar slabs of material comprised of Medium 1, Medium 2, and Medium 1, respectively, as depicted in Fig. \ref{fig_refocus}(a). 
From left to right, the layers have thicknesses $d_0 = 0.5\lambda_0, d_1 = \lambda_0, d_2 = 0.5\lambda_0$ where $\lambda_0  \approx 277.98$ nm, the wavelength in both media at $\omega = \omega_c$. 
The associated wavevector $k_0 = 2\pi/\lambda_0 $ is purely real. 

In this setup, a line source with a Gaussian spatial distribution with standard deviation $0.1 \lambda_0$ is placed in the leftmost layer of Medium 1 at $(x,z)=(1,0) \lambda_0$, as shown in Fig. \ref{fig_refocus}(b). The plot shows the magnitude of the Poynting vector when the source has complex frequency $\omega = \omega_c$.
%
We see that the field is refocused at the midpoint of layer 2 and again at the rightmost point of layer 3, achieving the double focusing effect of an ideal perfect lens \cite{pendry_prl_2000}. 
For visual clarity, the plot shows only propagating waves due to the exponential growth of the evanescent waves. 
The transmission coefficient through the three-layer configuration plotted as a function of the parallel wavevector $k_\parallel$ is shown in Fig. \ref{fig_refocus}(c). The red dashed lines mark the boundaries between propagating and evanescent waves, which occurs at $k_\parallel = \pm k_0$. We see that at complex frequency, the transmission is unity for both propagating and evanescent waves. Moreover, the phase remains unaffected, since $\text{Im}[t] = 0$.
As a comparison, at the real frequency $\omega = \omega_r$, no double focusing behavior is observed [Figs. \ref{fig_refocus}(d) and (e)], as expected since the two media have positive indices at $\omega = \omega_r$. 

Our results here indicate that the ideal perfect lens, which performs both refocusing of propagating waves and recovery of evanescent waves, can be achieved using passive lossy media operating at complex frequencies. We note that the results here differ from those in Ref. \cite{loss_compensation_alu_PRX_2023}, which focused on the recovery of evanescent field components, corresponding to the so-called “poor-man’s superlens,” as defined by Pendry \cite{pendry_prl_2000}. 

\subsubsection{Superscattering}
As a final example, we demonstrate the phenomenon of superscattering that is enabled by optical antimatter at complex frequency. Theoretical proposals for superscatterers hinge on the use of ideal negative-index materials \cite{superscatterer_pendry_2009, pendry2015transforming} or multiple overlapping resonances in plasmonic-dielectric-plasmonic layered structures \cite{zhichao_superscattering_PRL_2010, design_superscat_zhichao_fan_2011}. In both cases, the experimental realization of superscatterers, especially at optical frequencies, is hindered by the loss present in the required materials. Here, we show that the effects of superscattering can be achieved with passive, lossy materials at a complex frequency.

In Fig. \ref{fig_superscat}(a), we consider a cylinder of air surrounded by an annulus of Medium 2, embedded in Medium 1. The radius of the air cylinder and the thickness of the annulus are both $18.75$ nm, giving an overall radius of $r = 37.5$ nm. The white dashed line denotes the location of the TM-polarized plane wave source. 
Since Media 1 and 2 form a complementary pair, the annulus optically cancels a portion of the embedding medium when excited at the complex frequency $\omega = \omega_c$. 
The resulting field distribution is shown in Fig. \ref{fig_superscat}(b). Due to the optical antimatter effects of the annulus, the field distribution appears as if scattering off a larger object. 
We validate this by comparing to the scattering of an air cylinder with radius $R = 66.6$ nm, as depicted in Fig. \ref{fig_superscat}(c). 
The field distributions of the two scatterers are almost identical, as can be seen in Fig. \ref{fig_superscat}(b) and (d). The two scatterers have
the same scattering cross section $\sigma_{\text{scat}} = 0.2767$ $\mu$m. 
On the other hand, the structure in Fig. \ref{fig_superscat}(c) has a radius that is nearly twice that of the outer radius of the annulus in Fig. \ref{fig_superscat}(a). The results here illustrate the ability to use passive, lossy media to achieve a superscattering effect by operating at a complex frequency. 

\section{Discussion}


In summary, we have demonstrated the concept of optical antimatter using only \textit{passive, lossy} materials at complex frequency. We further established that one can
engineer arbitrary complex-valued permittivity and permeability in such materials. We find that \textit{positive-index} Lorentz-Drude materials can be harnessed to achieve effective \textit{negative-index} materials when excited at complex frequency. Furthermore, we have shown that such a concept can be applied to achieve ideal negative index lensing and superscattering, which have both been difficult to realize due to losses, especially at optical frequencies. In this work, we have highlighted the interplay between complex frequency and complex material parameters to achieve lossless propagation within the media and effective refractive index manipulation. Future directions include exploring other possibilities in transformation optics such as invisibility cloaking \cite{leonhardt_optical_conform_mapping_2006}, as well as non-Hermitian phenomena such as exceptional points and perfect absorbers \cite{absorbing_excep_points_absorbers_2019}. 
%
Our work points to avenues for using temporally structured light to realize exotic optical behaviors.

\section{Materials and Methods}
The electromagnetic field demonstrations of optical antimatter with complex frequency (Figs. \ref{optical_antimatter} and \ref{fig_superscat}) were computed using an open-source Finite Difference Frequency Domain (FDFD) code \cite{ceviche_ref_2019}. In Fig. \ref{optical_antimatter}, Bloch boundary conditions were applied to the transverse direction to impose the appropriate phase shift for the excitation of the off-normal incident light. Along the propagation direction, Perfectly Matched Layer (PML) boundary conditions were used to absorb outgoing waves. The existing code was modified to accommodate double negative media ($\epsilon,\mu<0$) and to account for the complex frequency in the PML media. In Fig. \ref{fig_superscat}, PML boundary conditions were applied along all directions to absorb the scattered waves.
Scattering cross sections for the superscatterer and air cylinder in Fig. 3 were calculated using the Total-Field/Scattered-Field (TFSF) framework implemented within the FDFD simulations \cite{TFSF_in_FDFD_2012}. The TFSF code was adapted to accommodate background media with arbitrary permittivity and permeability. Further calculation details can be found in the Supplementary Materials \cite{SM}.
Field plots in Fig. \ref{fig_refocus} were computed using a scattering-matrix code specifically written to ensure numerical stability at complex frequency excitations \cite{s-matrix_culshaw_1999}.

\begin{acknowledgments}
This work is supported by Samsung Electronics, and by a MURI grant from the U. S. Air Force Office of Scientific Research (Grant No. FA9550-21-1-0244). O. Y. L. is supported by a Stanford Graduate Fellowship.
\end{acknowledgments}

\bibliography{bibliography}

\ifarXiv
    

\section*{Supplementary material}

# Supplementary Material for: "Ideal optical antimatter using passive lossy materials under complex frequency excitation"

Olivia Y. Long,[1,2,*] Peter B. Catrysse,[2] Seunghoon Han,[3,4] and Shanhui Fan[1,2,5,†]

[1]*Department of Applied Physics, Stanford University, Stanford, California 94305, USA*
[2]*Edward L. Ginzton Laboratory, Stanford University, Stanford, California 94305, USA*
[3]*Samsung Advanced Institute of Technology, Samsung Electronics, Suwon-si 16678, South Korea*
[4]*Semiconductor R&D Center, Samsung Electronics, Hwaseong-si 18448, South Korea*
[5]*Department of Electrical Engineering, Stanford University, Stanford, California 94305, USA*
(Dated: May 30, 2025)

## CONTENTS

### I. DERIVATION OF MATRIX EXPRESSION IN MAIN TEXT

In this section, we derive the matrix $M$ and vector $\mathbf{b}$ given in Eq. 10 of the main text. We start with $\epsilon(\omega) = C + iD$ where $\omega = \omega' + i\omega''$:

$$\epsilon(\omega) = 1 + \frac{\omega_p^2}{A - iB} = C + iD \tag{S1}$$

Here, we have used $A \equiv -[\omega_0^2 - \omega'^2 + \omega''(\omega'' + \gamma)]$ and $B \equiv \omega'(2\omega'' + \gamma)$. Multiplying both sides of Eq. S1 by $A - iB$, we have:

$$A - iB + \omega_p^2 = (C + iD)(A - iB) \tag{S2}$$

Matching the real and imaginary parts, we get:

$$A + \omega_p^2 = AC + BD \tag{S3}$$
$$-B = AD - BC \tag{S4}$$

Expanding out the expressions for $A$ and $B$, the two equations become:

$$[\omega'^2 - \omega_0^2 - \omega''(\omega'' + \gamma)] + \omega_p^2 = C[\omega'^2 - \omega_0^2 - \omega''(\omega'')] + \omega'(2\omega'' + \gamma)D \tag{S5}$$
$$-\omega'(2\omega'' + \gamma) = D[\omega'^2 - \omega_0^2 - \omega''(\omega'' + \gamma)] - \omega'(2\omega'' + \gamma)C \tag{S6}$$

---

[*] olong@stanford.edu
[†] shanhui@stanford.edu

Combining like terms and writing in matrix form, we arrive at the matrix expression in Eq. 10–13 of the main text:

$$\begin{bmatrix} C-1 & 1 & \omega''(C-1) - \omega'D \\ D & 0 & \omega'(C-1) + \omega''D \end{bmatrix} \begin{bmatrix} \omega_0^2 \\ \omega_p^2 \\ \gamma \end{bmatrix} = \begin{bmatrix} (C-1)(\omega'^2 - \omega''^2) + 2\omega'\omega''D \\ D(\omega'^2 - \omega''^2) - 2\omega'\omega''(C-1) \end{bmatrix} \quad (S7)$$

## II. DERIVATION OF $\alpha$ INTERVAL OVER WHICH x > 0

In this section, we derive the range of $\alpha$ in which the components of **x** are all positive, where $\alpha$ is the free variable from Eq. 15 of the main text. The derivation here provides additional information beyond Theorem 1 in the main text, which states that:

$$\exists \alpha : \quad x_k(\alpha) > 0 \quad \forall k. \quad (S8)$$

Note that the component $(x_n)_2 > 0$ since $\omega' > 0$. Thus, for all $\alpha > 0$, $x_2(\alpha) > 0$ since $(x_p)_2 = 0$.

For $k \neq 2$, $(x_n)_k$ may be $> 0$ or $< 0$ depending on the values of $C$ and $D$. If $(x_n)_k > 0$, $x_k(\alpha) > 0$ when:

$$\alpha > -\frac{(x_p)_k}{(x_n)_k} \quad (S9)$$

If $(x_n)_k < 0$, $x_k(\alpha) > 0$ when:

$$\alpha < -\frac{(x_p)_k}{(x_n)_k} \quad (S10)$$

Hence, the values of $\alpha$ satisfying Eq. (S8) lie in the interval:

$$\alpha \in \left( \max_{k \in I^+} \left( -\frac{(x_p)_k}{(x_n)_k} \right), \min_{k \in I^-} \left( -\frac{(x_p)_k}{(x_n)_k} \right) \right) \quad (S11)$$

where $I^+ = \{ k \in \{1,3\} \mid (x_n)_k > 0 \}$ and $I^- = \{ k \in \{1,3\} \mid (x_n)_k < 0 \}$.

Since $(x_p)_k > 0$ for $k \neq 2$, the interval in Eq. (S11) has a negative lower bound and a positive upper bound, which guarantees that it is non-empty. Thus, there exists $\alpha > 0$ in the interval of Eq. (S11), satisfying $x_2(\alpha) > 0$ and completing the proof of Eq. (S8). In other words, we have shown that Eqns. 10–13 in the main text can be satisfied by a passive, lossy material for arbitrary $C, D \in \mathbb{R}$.

## III. CALCULATION OF NULL-SPACE VECTOR

In this section, we calculate the cross product of the vectors $\mathbf{r}_1, \mathbf{r}_2$ to find a nullspace vector $\mathbf{x}_n$ (Eq. 16 of main text):

$$\mathbf{x}_n = \mathbf{r}_1 \times \mathbf{r}_2 = \begin{vmatrix} \mathbf{i} & \mathbf{j} & \mathbf{k} \\ C-1 & 1 & \omega''(C-1) - \omega'D \\ D & 0 & \omega'(C-1) + \omega''D \end{vmatrix}$$

$$= [\omega'(C-1) + \omega''D]\mathbf{i} - \omega'[D^2 + (C-1)^2]\mathbf{j} - D\mathbf{k}. \quad (S12)$$

## IV. ELECTROMAGNETIC FIELD DEMONSTRATIONS FOR DIFFERENT GEOMETRIES AT COMPLEX FREQUENCY

In Fig. S1, we demonstrate optical antimatter using another structure comprised of a slab with planar inhomogeneity along the transverse direction. In panel (a), a layer of Medium 3 (yellow) with thickness 0.6 $\mu$m along the transverse direction is embedded in Medium 1 (blue), situated in a background of air. The slab has thickness 0.6 $\mu$m along the

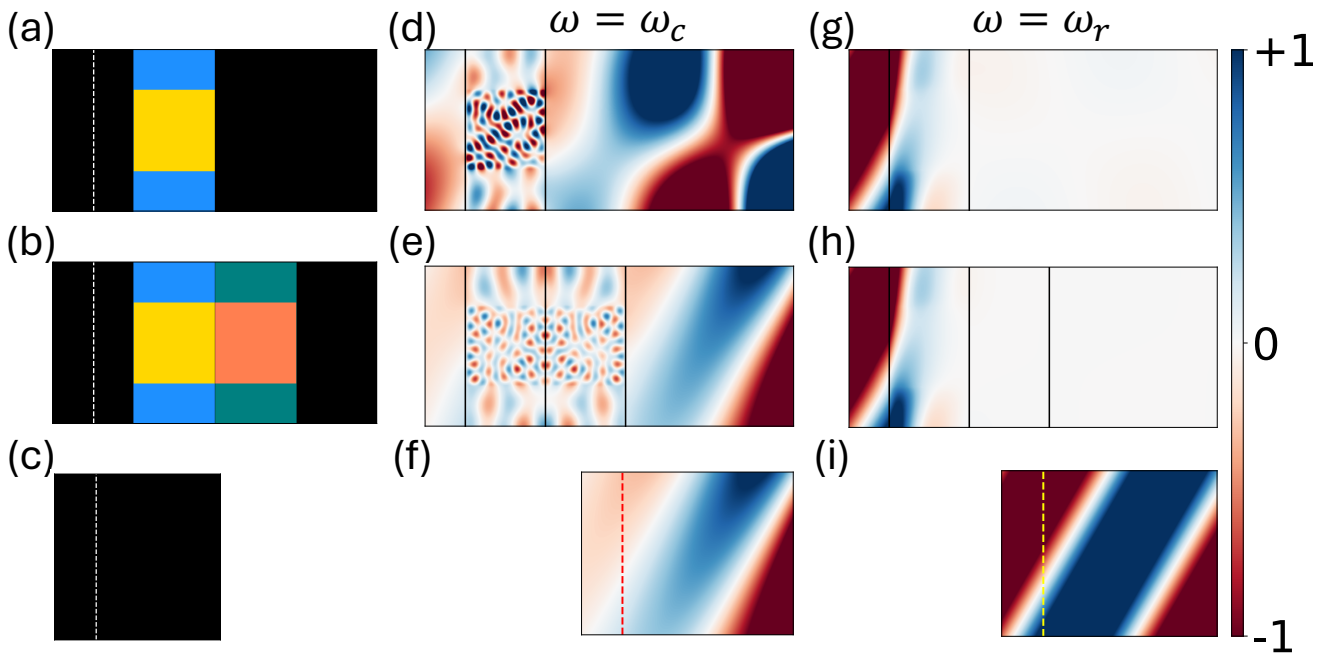

FIG. S1. Optical antimatter for planar inhomogeneity along the transverse direction. In panels (a)–(c), the colors blue, teal, yellow and orange correspond to Media 1, 2, 3, 4, as described in Table I of the main text, respectively. Black denotes vacuum. (a) Schematic of structure with a layer of Medium 3 (yellow) with thickness 0.6 $\mu$m along the transverse direction is embedded in Medium 1 (blue), situated in a background of air. The slab has thickness 0.6 $\mu$m along the propagation direction. Dashed white line indicates location of TM-polarized plane wave source. (b) Schematic of structure placed adjacent to its complementary counterpart. (c) Free space configuration with plane wave source. (d)–(f) TM-polarized Re[$H_z$] field patterns at the complex frequency $\omega = \omega_c$ for the configurations in (a), (b), and (c), respectively. (g)–(i) TM-polarized Re[$H_z$] field patterns at the real frequency $\omega = \omega_r$ for the configurations in (a), (b), and (c), respectively. In (f) and (i), the red and yellow dashed lines mark the effective location of the optically cancelled complementary media pair, respectively. For each $\omega$, field plots are normalized to the same maximum and minimum values. All field plots were generated using FDFD [40], with Bloch boundary conditions in the transverse ($y$) direction and PML boundary conditions in the propagation ($x$) direction. Plots show 0th diffraction order in free space.

propagation direction. The top right panel shows that the fields transmitted through the structure strongly deviate from the incident plane wave. By placing the complementary slab structure adjacent to the single slab, we are once again able to add optical antimatter to optically cancel the scattering from the first slab. As shown in panel (e), we see that the transmitted fields emerging to the right of the complementary pair match the free-space case in Fig. S1(f).

In Fig. S2(a)–(b), we demonstrate the field pattern recovery of a scatterer placed adjacent to a pair of complementary media. In the top panel of (a), we show the schematic of the setup. The top panel of (b) shows the resulting field pattern at complex frequency $\omega = \omega_c$. We can compare the field pattern to that of just the scatterer shown schematically in the lower panel of (a), and we see that the emerging fields on the right hand side are the same [lower panel of (b)]. In Figs. S2(c)–(d), we demonstrate the optical antimatter of two nested sets of complementary media, which cancel pairwise, as imagined in the original proposal [1]. The configuration is shown schematically in the top panel of (c). The inner two slabs form a complementary pair and effectively cancel each other, while the outer two slabs cancel. As a result, the fields on the right hand side of the structure emerge as if unperturbed, as seen in the top panel of (d). We can compare the fields to those of the incident plane wave propagating in free-space, shown schematically in the lower panel of (c), and we see that the fields on the right side of the dashed red line in the lower panel of (d) match those on the right side of the slab structure [top panel of (d)].

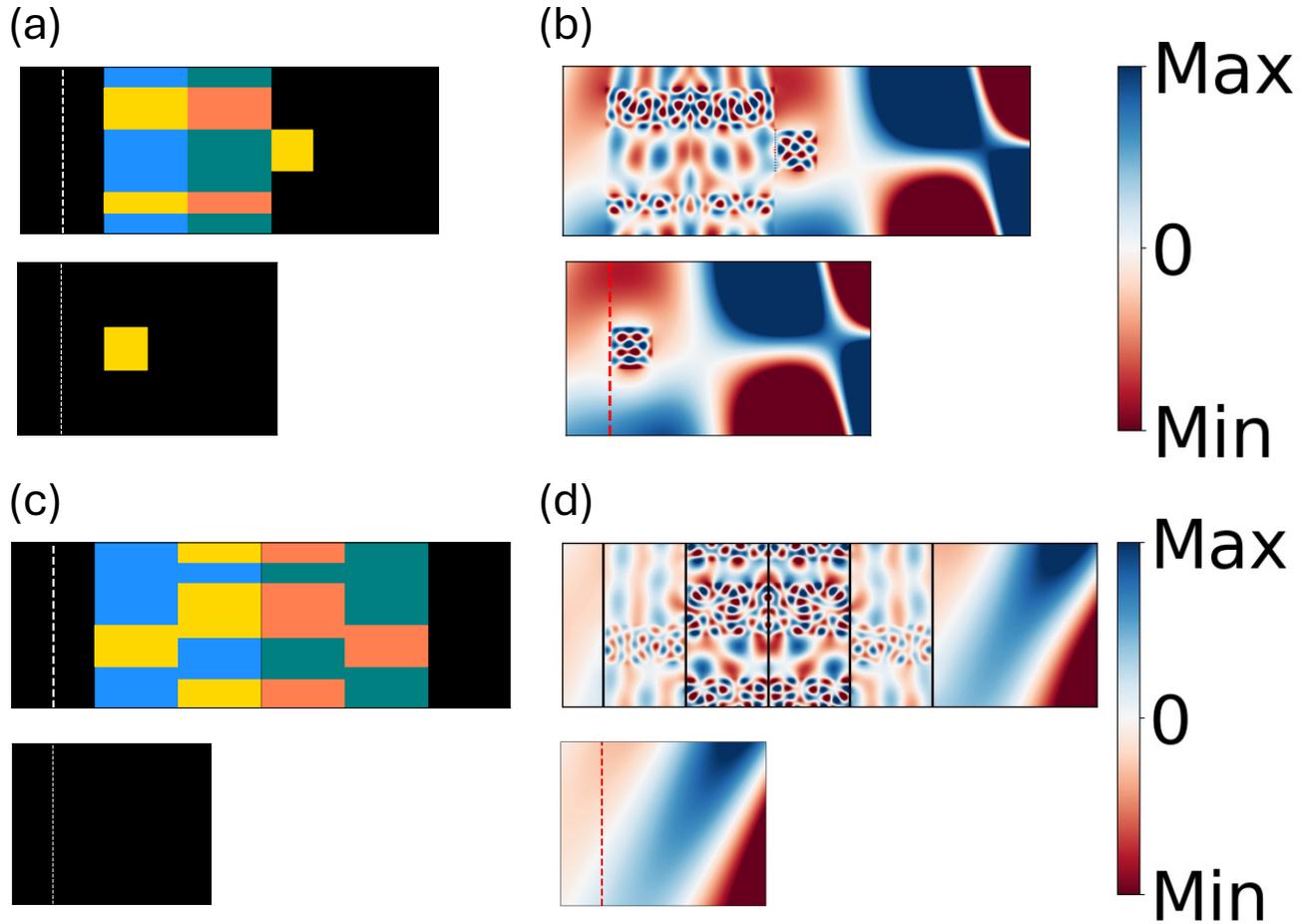

FIG. S2. Field pattern recovery of a scatterer using optical antimatter. (a) Schematic of complementary pair placed in front of scatterer in free space (top panel) and equivalent configuration of the scatterer in free space, with the space occupied by the complementary pair removed (bottom panel). (b) TM-polarized Re[$H_z$] field patterns at the complex frequency $\omega = \omega_c$ for the configurations shown in (a). (c) Nested configuration of two pairs of complementary media (top panel) and equivalent configuration in free space (bottom panel). The two innermost layers and the two outermost layers for complementary pair. Taken pairwise, the overall scattering pattern is cancelled. (d) TM-polarized Re[$H_z$] field patterns at the complex frequency $\omega = \omega_c$ for the configurations shown in (c). In all panels showing schematics, the dotted white line indicates plane wave source location. The dotted red lines in the field patterns indicate effective location of the optical antimatter structure that is rendered invisible.

## V. TOTAL-FIELD/SCATTERED-FIELD SIMULATION DETAILS

Fig. S3 shows the setup for computing the scattering cross section using the TFSF framework within our FDFD simulations. The scattering cross sections of the structures shown in Fig. 3 of the main text were computed by taking the line integral of the Poynting vector over the bounding box (depicted in red) outside the TFSF box and normalizing by the incident power $S_x$ inside box with no scatterer. The TFSF code was modified to accommodate for background media with arbitrary permittivity and permeability values.

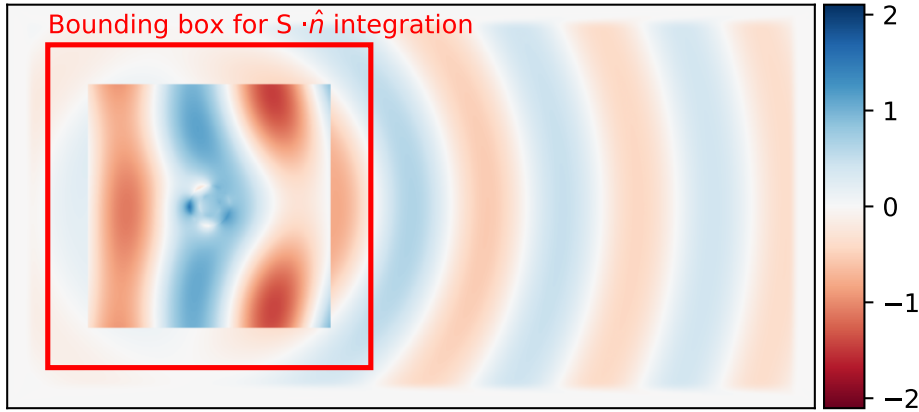

FIG. S3. Schematic of Total Field Scattered Field setup in FDFD simulations. The outgoing Poynting flux is integrated along the bounding box shown in red.

## VI. VIRTUAL GAIN IN LORENTZ-DRUDE MATERIAL AT COMPLEX FREQUENCY

In this section, we derive the conditions for achieving $\text{Im}[\epsilon] < 0$ (virtual gain) in a Lorentz-Drude material, without using intrinsic material gain. At a given complex frequency $\omega = \omega' + i\omega''$, we have:

$$\epsilon(\omega) = 1 + \frac{\omega_p^2}{\omega_0^2 - \omega^2 - i\omega\gamma} = 1 + \frac{\omega_p^2}{(\omega_0^2 - \omega'^2 + \omega''^2 + \omega''\gamma) - i\omega'(2\omega'' + \gamma)} \tag{S13}$$

$$= 1 + \frac{\omega_p^2}{A - iB} = \frac{[A(A + \omega_p^2) + B^2] + iB\omega_p^2}{A^2 + B^2} \tag{S14}$$

where $A, B$ are defined in Section I above. To achieve $\text{Im}[\epsilon] < 0$, the condition is:

$$B\omega_p^2 < 0 \tag{S15}$$
$$\omega_p^2 \omega'(2\omega'' + \gamma) < 0 \tag{S16}$$

Since we would like to have a material with no intrinsic gain (i.e. $\omega_p^2 > 0$), the condition becomes:

$$\omega'(2\omega'' + \gamma) < 0 \tag{S17}$$

Since $\omega' > 0$, this gives us the condition on $\gamma$:

$$\gamma < -2\omega'' \tag{S18}$$

Thus, for $\gamma > 0$ as must be true in a physical Lorentz-Drude material, we must have $\omega'' < 0$, which corresponds to a temporally decaying wave. This is also the reason we choose to work with decaying waves rather than growing waves in time to achieve complementary media and optical antimatter effects.


\fi

\end{document}